\documentclass[final,3p,times]{elsarticle}
\usepackage{amssymb}
\usepackage{amsmath}
\begin{document}

\begin{frontmatter}



\title{Charmed baryon decays at BESIII}


\author{Xudong Yu On behalf of the BESIII Collaboration} %% Author name
\ead{yuxd@stu.pku.edu.cn}

\affiliation{organization={School of Physics,Peking University},%Department and Organization
            addressline={209 Chengfu Road}, 
            city={Beijing},
            postcode={100871}, 
            country={People’s Republic of China}}

%% Abstract
\begin{abstract}
BESIII has accumulated 4.5~fb$^{-1}$ of $e^+e^-$ collision data in the 4.6 to 4.7~GeV energy range, corresponding to the world's largest sample of $\Lambda_c^+\bar{\Lambda}_c^-$ pairs. This paper summarizes recent BESIII results on charmed-baryon decays, including the observation of the rare semi-leptonic decay $\Lambda_c^+\to ne^+\nu_e$ using a Graph Neural Network, the first measurement of the decay asymmetry in the pure $W$-exchange decay $\Lambda_c^+\to\Xi^0K^+$, and branching fraction measurements of the inclusive decays $\Lambda_c^+\to Xe^+\nu_e$ and $\bar{\Lambda}_c^-\to \bar{n}X$. We also report partial wave analyses of $\Lambda_c^+\to\Lambda\pi^+\pi^0$ and $\Lambda_c^+\to\Lambda\pi^+\eta$, measurements of Cabibbo-suppressed decays such as $\Lambda_c^+\to p\pi^0$, and studies of $K_S^0-K_L^0$ asymmetries in $\Lambda_c^+$ decays.
\end{abstract}



%% Keywords

\end{frontmatter}



\section{Introduction}\label{sec:intro}

$\Lambda_c^+$ is the lightest charmed baryon. Since most bottom baryons and excited charmed baryons eventually decay to $\Lambda_c^+$, it plays a central role in heavy-flavor spectroscopy and decay studies. Located in the transition region between perturbative and non-perturbative QCD, $\Lambda_c^+$ decays receive sizable non-perturbative contributions. In contrast to charm-meson decays, non-factorizable $W$-exchange amplitudes in $\Lambda_c^+$ decays are neither color suppressed nor helicity suppressed, and can therefore be comparable to, or even larger than, factorizable contributions. Their theoretical evaluation remains challenging, and a variety of phenomenological approaches have been developed, including the constituent quark model, MIT bag model, pole model, and current algebra in the soft-pion limit. More recently, $S\!U(3)$ flavor symmetry has been widely used to relate different topological diagram amplitudes or irreducible representation amplitudes among charmed-baryon decay modes. Global fits can be performed using existing experimental data to determine the common amplitudes and thereby predict the branching fractions of yet-unobserved decay channels. Measurements of $\Lambda_c^+$ decays therefore provide essential tests of these approaches and important inputs to theory.

The BESIII detector records symmetric $e^+e^-$ collisions delivered by the BEPCII storage ring, which operates with a peak luminosity of $1\times10^{33}\,\mathrm{cm}^{-2}\mathrm{s}^{-1}$ in the center-of-mass energy range from 2.0 to 4.95~GeV. Benefiting from the excellent detector performance, BESIII has collected 4.5~fb$^{-1}$ of data between 4.6 and 4.7~GeV, together with an additional 1.9~fb$^{-1}$ between 4.74 and 4.95~GeV. These samples enable precision studies of charmed baryons using both the single-tag (ST) and double-tag (DT) techniques.

\section{Inclusive decays}\label{sec:inclusive}

The absolute branching fractions of the inclusive decays $\Lambda_c^+\to Xe^+\nu_e$, $\bar{\Lambda}_c^-\to \bar{n}X$, and $\Lambda_c^+\to K_S^0 X$ are measured to be $(4.06 \pm 0.10_{\mathrm{stat}} \pm 0.09_{\mathrm{syst.}})\%$~\cite{BESIII:2022cmg}, $(32.4 \pm 0.7_{\mathrm{stat}} \pm 1.5_{\mathrm{syst.}})\%$~\cite{BESIII:2022onh}, and $(10.9 \pm 0.2 \pm 0.1)\%$~\cite{BESIII:2025cel}, respectively, where $X$ denotes any allowed particle system. The precision of $\mathcal{B}(\Lambda_c^+\to Xe^+\nu_e)$ is improved by more than a factor of three, and, together with $\mathcal{B}(\Lambda_c^+\to \Lambda e^+\nu_e)$, implies that the remaining semi-leptonic branching fraction is only of order $10^{-3}$. The ratio $\Gamma(\Lambda_c^+ \to Xe^+\nu_e) / \Gamma(D\to Xe^+\nu_e)=1.28\pm0.05$ agrees with the heavy-quark expansion prediction of 1.2, but disfavors the effective-quark estimate of 1.67~\cite{Manohar:1993qn,Gronau:2010if,Rosner:2012gj}. For $\bar{\Lambda}_c^-\to\bar{n}X$, a data-driven method~\cite{Liu:2021rrx} is used to model the anti-neutron response. Assuming negligible $CP$ asymmetry, our result improves the precision to about 5\% and suggests that roughly one quarter of neutron final states remain unobserved. The precision of $\mathcal{B}(\Lambda_c^+\to K_S^0 X)$ is also improved by a factor of three. Compared with the summed branching fraction of known exclusive modes, the result indicates a remaining contribution of $(3.0 \pm 0.4)\%$, consistent with the statistical isospin model~\cite{Gronau:2018vei}.

\section{Observation of semi-leptonic decay $\Lambda_c^+\to ne^+\nu_e$}\label{sec:nenu}

\begin{figure}[htbp]
    \centering
    \includegraphics[width=0.6\linewidth]{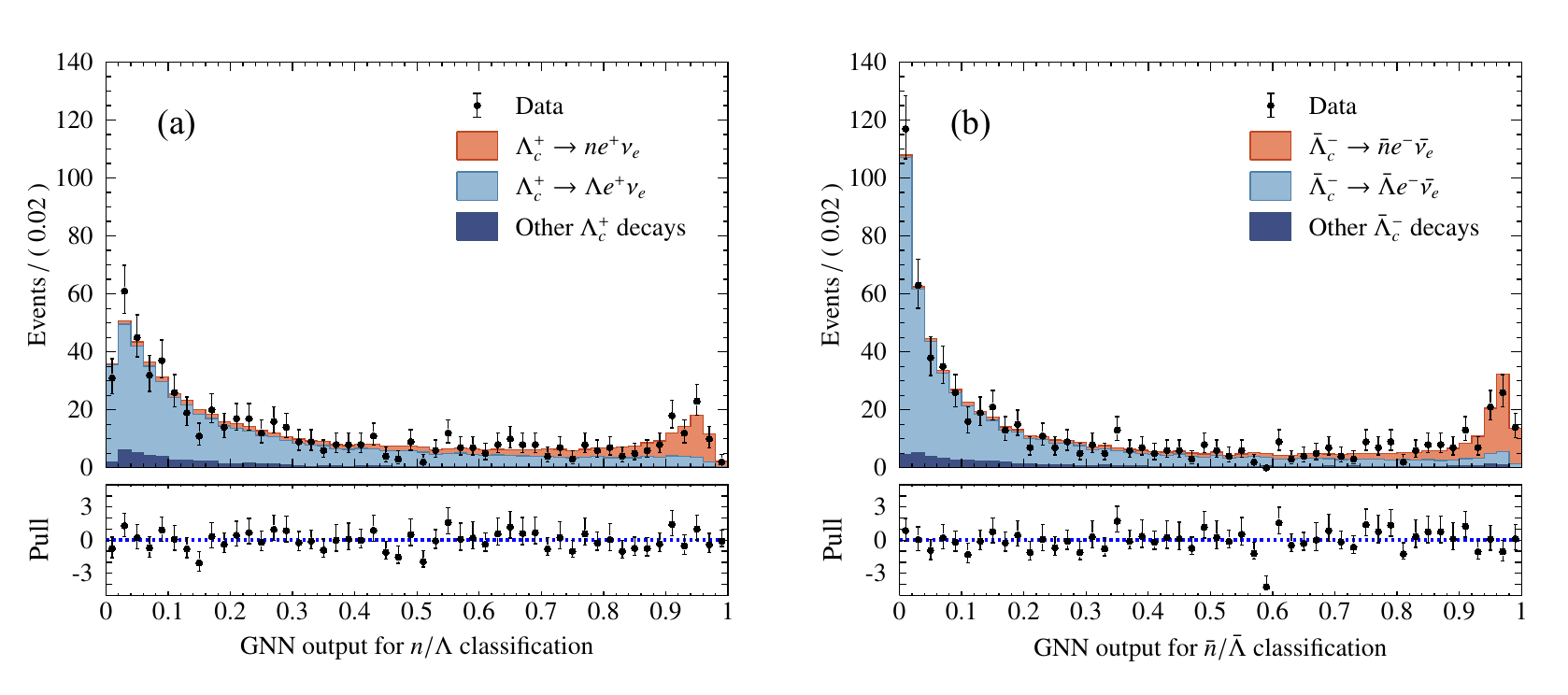}
    \includegraphics[width=0.33\linewidth]{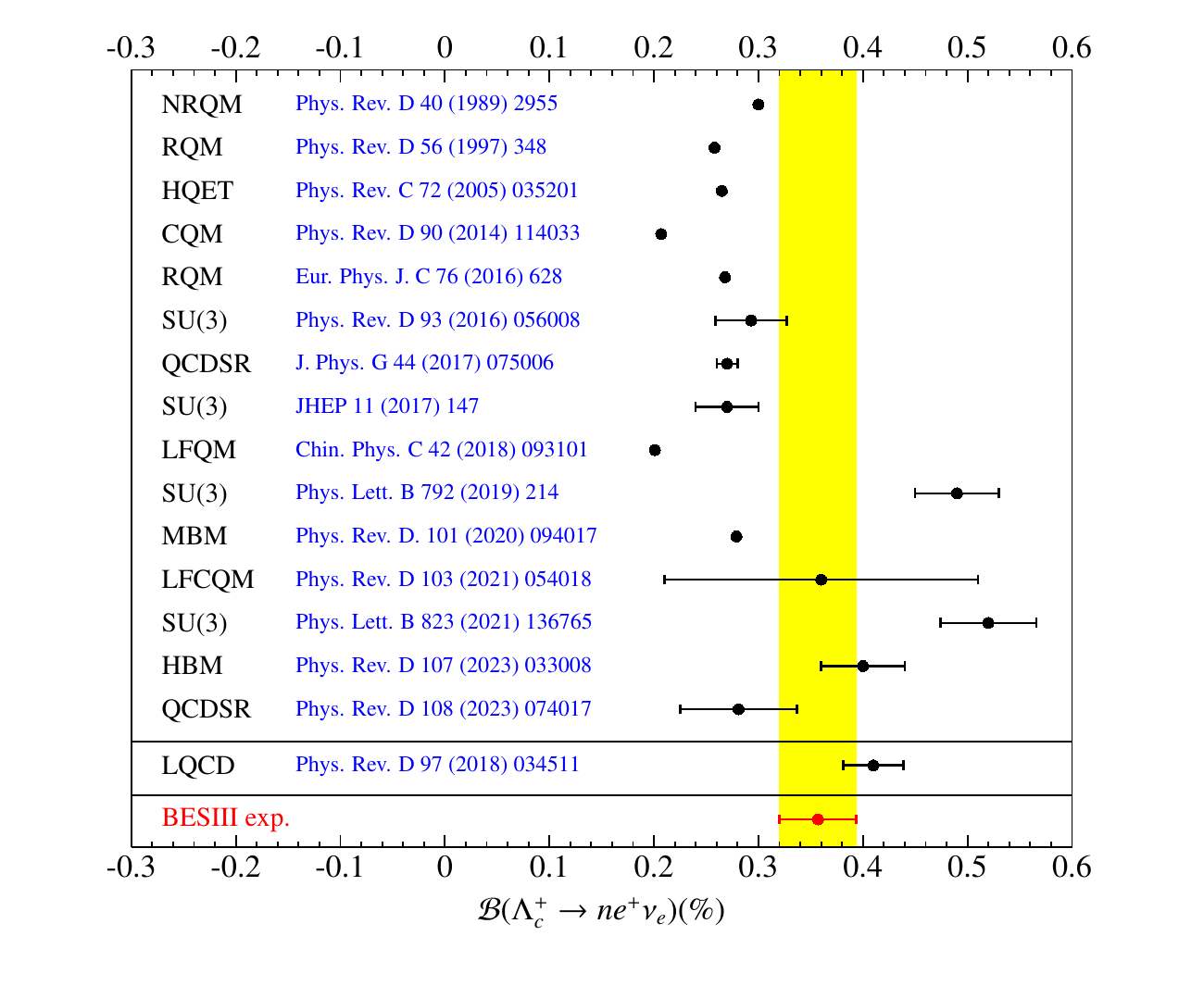}
    \caption{Left and middle: GNN-output distributions for $\Lambda_c^+\to n(\Lambda)e^+\nu_e$ and the charge-conjugate processes in data. Right: comparison of the BESIII result with theoretical predictions.}
    \label{fig:nenu}
\end{figure}

The semi-leptonic decay $\Lambda_c^+\to ne^+\nu_e$ is observed for the first time using a novel deep-learning approach~\cite{BESIII:2024mgg}. The dominant background comes from the Cabibbo-favored decay $\Lambda_c^+\to\Lambda e^+\nu_e$, because $\Lambda(\to n\pi^0)$ and a prompt neutron are difficult to distinguish with traditional methods. In this analysis, a Graph Neural Network (GNN) is used to classify the energy-deposition patterns in the electromagnetic calorimeter. Since neutron simulation does not perfectly reproduce data, a data-driven pipeline is established for GNN training, calibration, validation, and systematic uncertainty evaluation. As shown in Fig.~\ref{fig:nenu}, the corrected GNN-score distribution is used to extract the signal yield. The branching fraction is measured to be $(0.358\pm0.334_{\mathrm{stat.}}\pm0.014_{\mathrm{syst.}})\%$, with a statistical significance greater than $10\sigma$. Using the lattice-QCD form factors~\cite{Meinel:2017ggx} and the $\Lambda_c^+$ lifetime~\cite{Belle-II:2022ggx} as inputs, the CKM matrix element $|V_{cd}|$ is determined for the first time from charmed-baryon decays to be $0.208 \pm 0.011_{\mathrm{exp.}} \pm 0.007_{\mathrm{LQCD}} \pm 0.011_{\tau_{\Lambda_c^+}}$.

\section{Observation of the singly Cabibbo-suppressed decay $\Lambda_c^+\to p\pi^0$}\label{sec:ppi0}

\begin{figure}[htbp]
    \centering
    \includegraphics[width=0.28\linewidth]{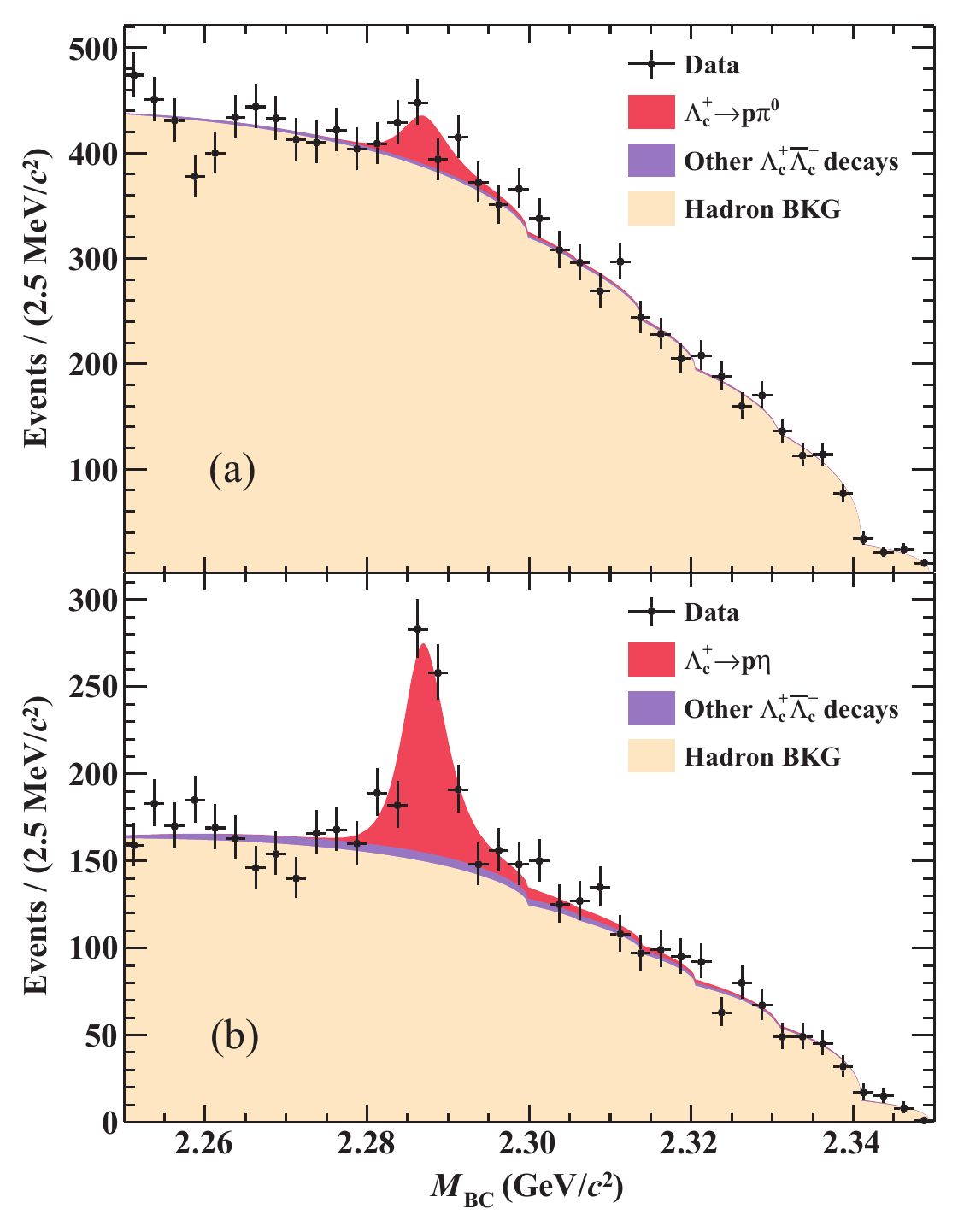}
    \includegraphics[width=0.28\linewidth]{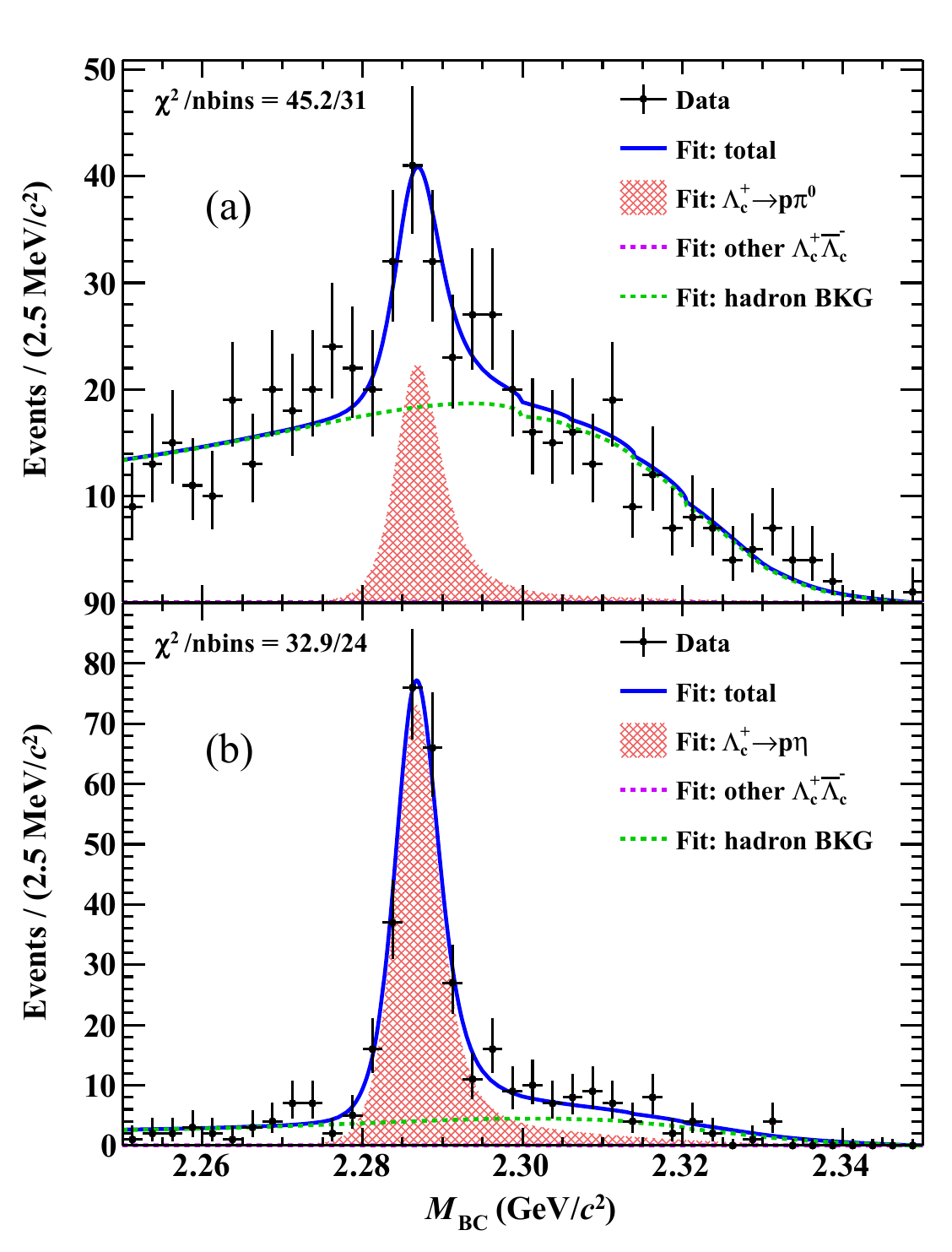}
    \includegraphics[width=0.43\linewidth]{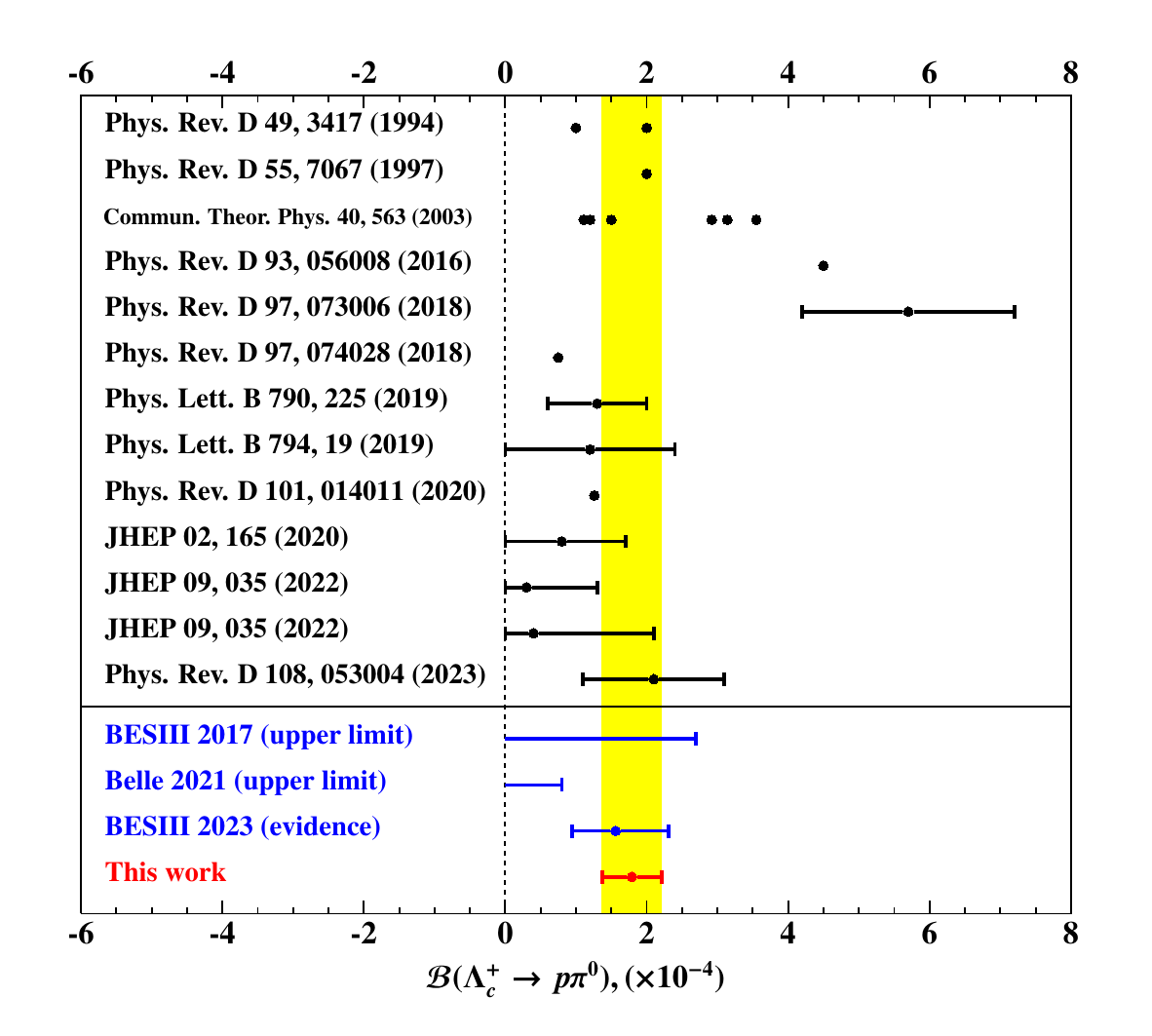}
    \caption{Left: $M_{\mathrm{BC}}$ distributions for $\Lambda_c^+\to p\pi^0$ and $\Lambda_c^+\to p\eta$ before the DNN selection. Middle: the corresponding distributions after the DNN selection. Right: comparison of the measured branching fraction with theoretical predictions.}
    \label{fig:ppi0}
\end{figure}

The singly Cabibbo-suppressed (SCS) decay $\Lambda_c^+\to p\pi^0$ has previously been studied by Belle~\cite{Belle:2021mvw} and BESIII~\cite{BESIII:2023uvs}, but the two results are in tension. Belle reported $\mathcal{B}(\Lambda_c^+\to p\pi^0) < 8.0\times10^{-5}$, while BESIII measured $\mathcal{B}(\Lambda_c^+\to p\pi^0) = (1.56^{+0.72}_{-0.58}\pm0.20)\times10^{-4}$ with a statistical significance of $3.7\sigma$. A more decisive measurement is therefore needed. To increase the signal yield, the ST method is adopted, providing about an order-of-magnitude gain in statistics compared with the DT approach. To suppress substantial hadronic backgrounds, a deep neural-network (DNN) signal--background classifier is employed. Low-level detector information, represented as a point cloud, is fed into the Particle Transformer (ParT) architecture. The reference channel $\Lambda_c^+\to p\eta$ is included in the DNN training to improve generalization and reduce systematic effects. As shown in Fig.~\ref{fig:ppi0}, the background is significantly suppressed after the DNN selection. The branching-fraction ratio is measured to be $\mathcal{B}(\Lambda_c^+\to p\pi^0) / \mathcal{B}(\Lambda_c^+\to p\eta) = 0.120\pm0.026$. Using the world-average value of $\mathcal{B}(\Lambda_c^+\to p\eta)$ as input, the absolute branching fraction is determined to be $(1.79\pm0.39_{\mathrm{stat.}}\pm0.11_{\mathrm{syst.}}\pm0.08_{\mathrm{ref.}})\times10^{-4}$~\cite{BESIII:2024cbr}.

\section{Measurements of $K_S^0-K_L^0$ asymmetries in $\Lambda_c^+$ decays}\label{sec:kskl}

The first measurements of $K_S^0-K_L^0$ asymmetries in charmed-baryon decays are performed using the channels $\Lambda_c^+\to pK_{L,S}^0$, $\Lambda_c^+\to pK_{L,S}^0\pi^+\pi^-$, and $\Lambda_c^+\to pK_{L,S}^0\pi^0$. The measured asymmetries are $\mathcal{R}(\Lambda_c^+\to pK_{L,S}^0) = -0.025\pm0.031$, $\mathcal{R}(\Lambda_c^+\to pK_{L,S}^0\pi^+\pi^-) = -0.027\pm0.048$, and $\mathcal{R}(\Lambda_c^+\to pK_{L,S}^0\pi^0) = -0.015\pm0.046$~\cite{BESIII:2024sfz}. These measurements provide useful input for searches for doubly Cabibbo-suppressed amplitudes.

\section{Decay asymmetry measurement of $\Lambda_c^+\to\Xi^0K^+$}\label{sec:xi0k}

The decay $\Lambda_c^+\to\Xi^0K^+$ proceeds purely through $W$ exchange. A long-standing challenge is to describe its branching fraction and decay asymmetry parameter $\alpha$ simultaneously within the same theoretical framework. In this work~\cite{BESIII:2023wrw}, an angular analysis is performed to extract the decay parameters: $\alpha = 0.01 \pm 0.16_{\mathrm{stat.}} \pm 0.03_{\mathrm{syst.}}$, $\Delta = 3.84 \pm 0.90_{\mathrm{stat.}} \pm 0.17_{\mathrm{syst.}}$, $\beta = -0.64 \pm 0.69_{\mathrm{stat.}} \pm 0.13_{\mathrm{syst.}}$, and $\gamma = -0.77 \pm 0.58_{\mathrm{stat.}} \pm 0.11_{\mathrm{syst.}}$. The quantity $\cos(\delta_p - \delta_s)$ is found to be close to zero, leading to two solutions for the phase difference: $\delta_p - \delta_s = -1.55 \pm 0.25_{\mathrm{stat.}} \pm 0.05_{\mathrm{syst.}}\,\mathrm{rad}$ and $\delta_p - \delta_s = 1.59 \pm 0.25_{\mathrm{stat.}} \pm 0.05_{\mathrm{syst.}}\,\mathrm{rad}$. This feature has not been considered in previous theoretical studies.

\begin{figure}[htbp]
    \centering
    \includegraphics[width=0.4\linewidth]{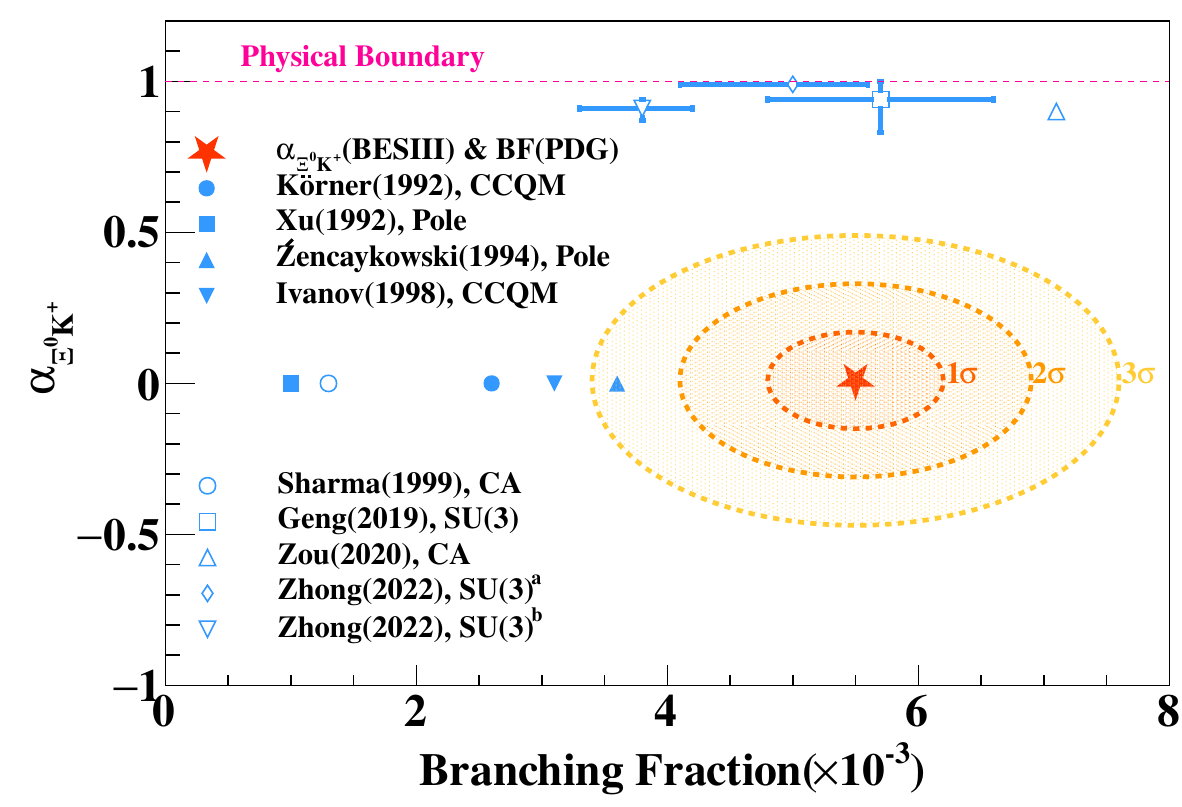}
    \caption{Comparison of experimental results for branching fraction and decay asymmetry of $\Lambda_c^+\to\Xi^0K^+$ decay with theoretical predictions.}
    \label{fig:xi0k}
\end{figure}

\section{Partial wave analysis of $\Lambda_c^+\to\Lambda\pi^+\pi^0$}\label{sec:pwa_lmdpipi0}

\begin{figure}[h]
    \centering
    \includegraphics[width=0.33\linewidth]{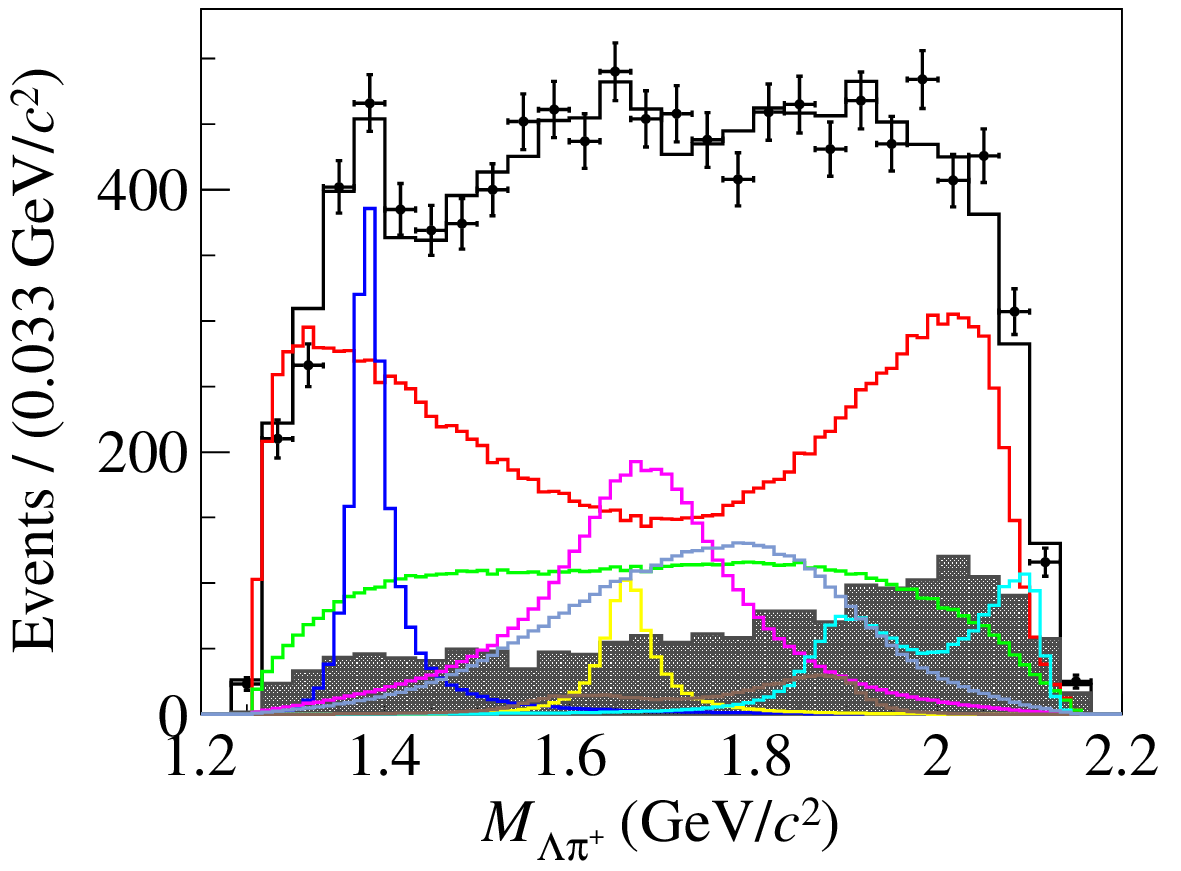}
    \includegraphics[width=0.33\linewidth]{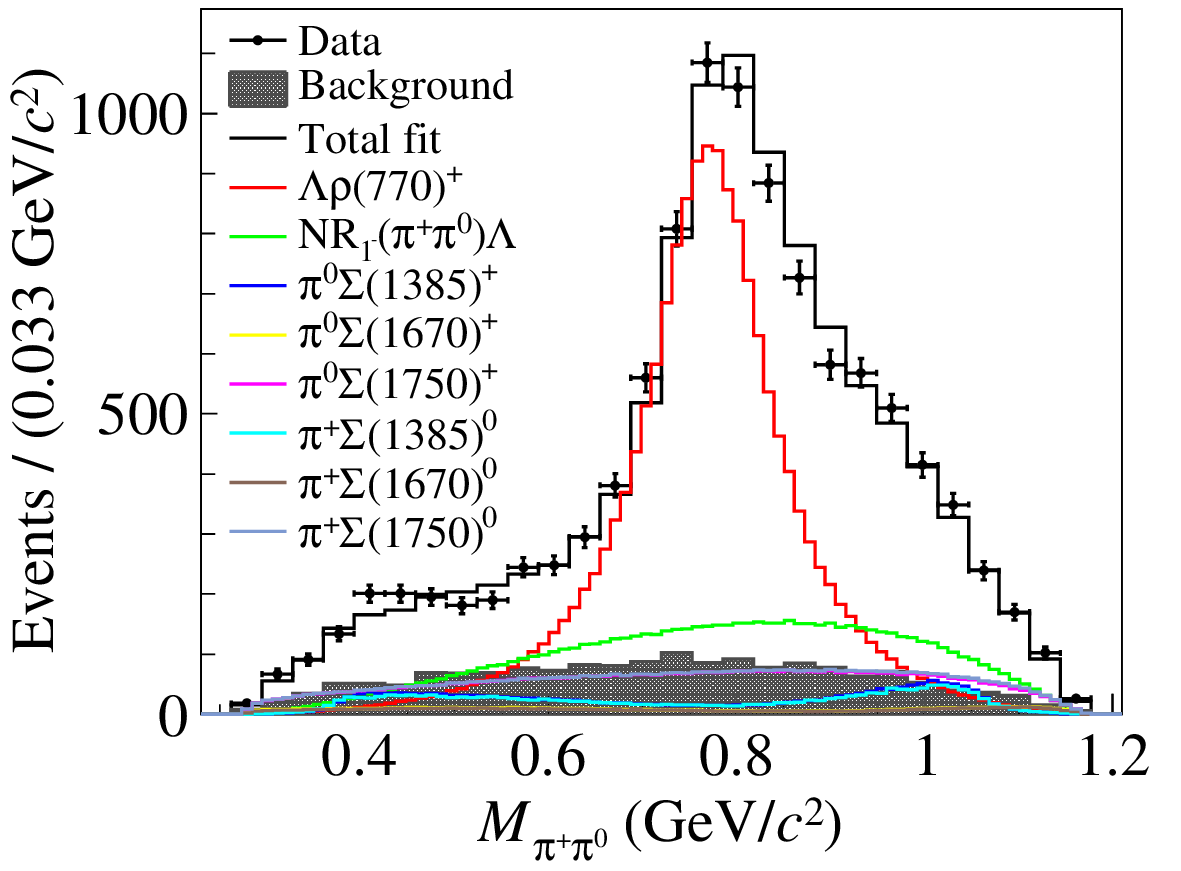}
    \includegraphics[width=0.33\linewidth]{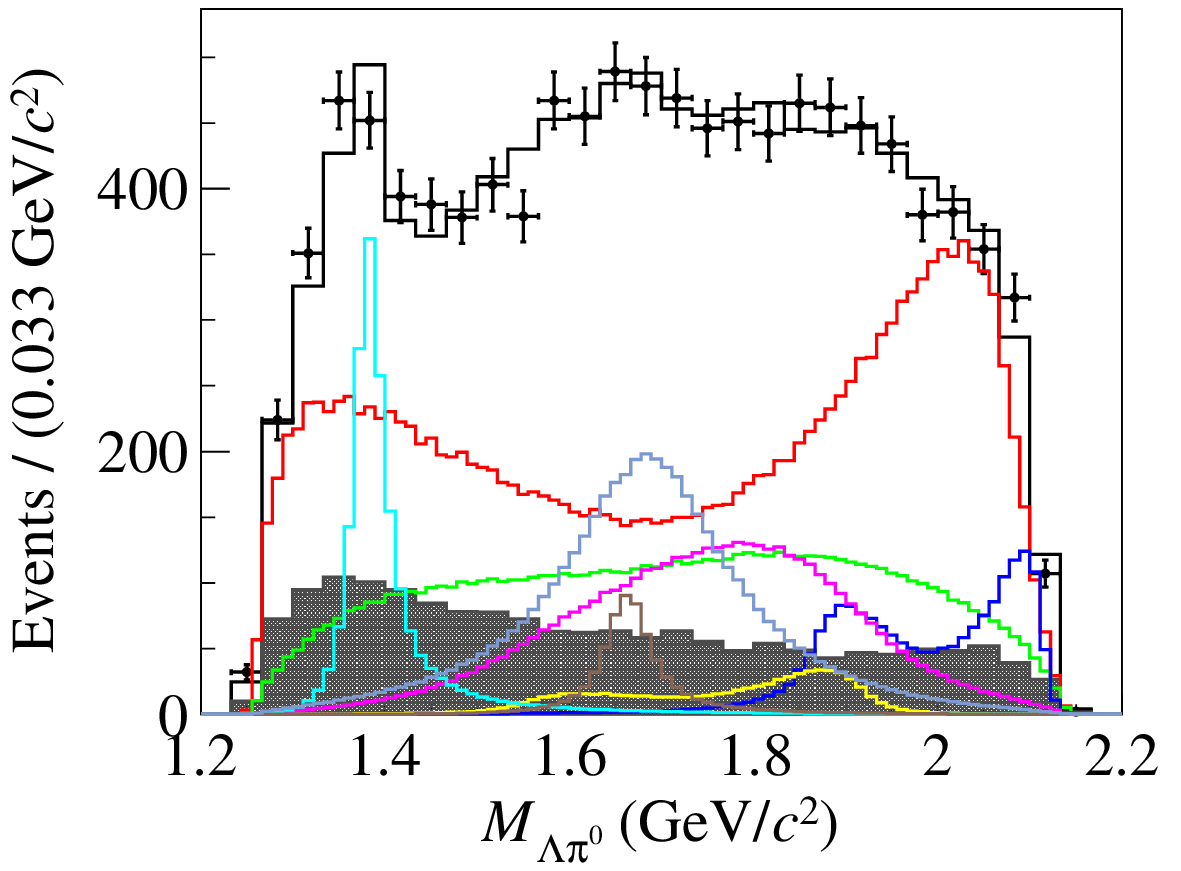}
    \includegraphics[width=0.33\linewidth]{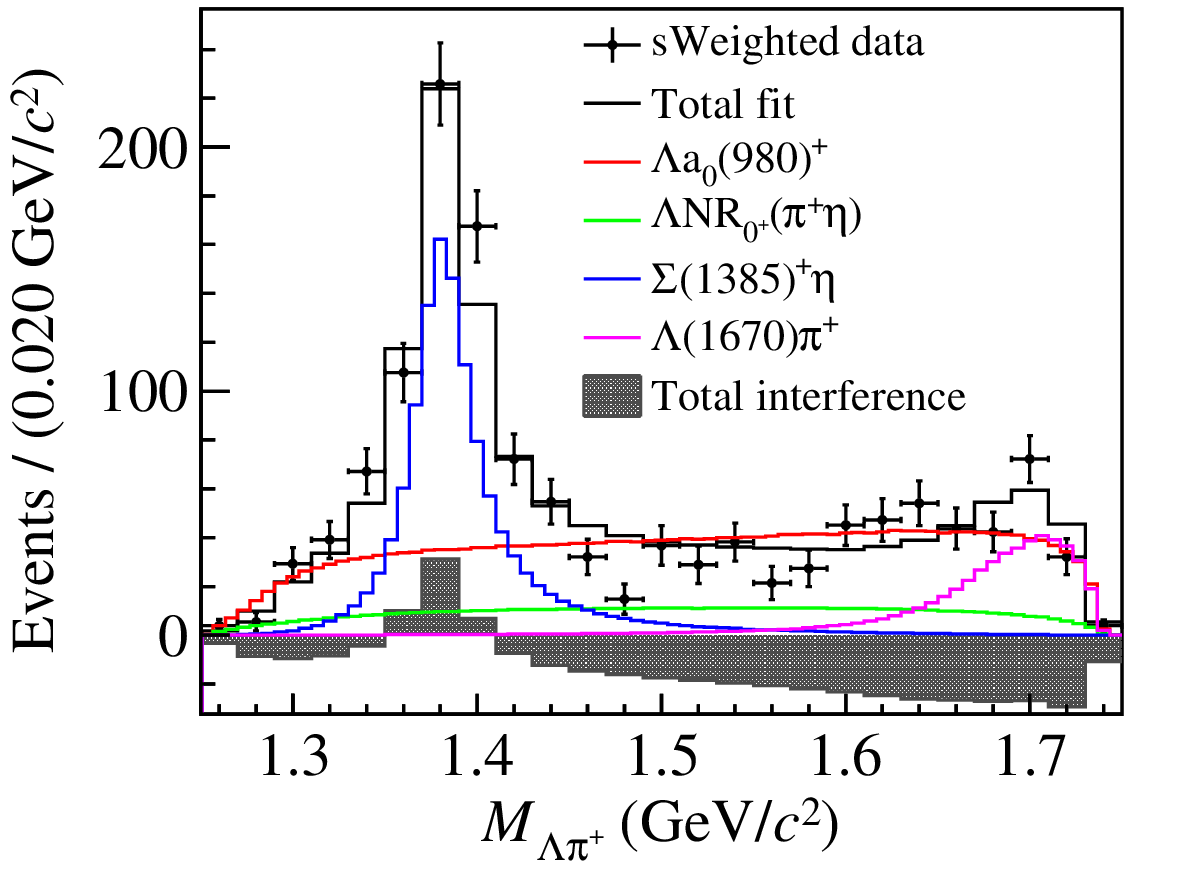}
    \includegraphics[width=0.33\linewidth]{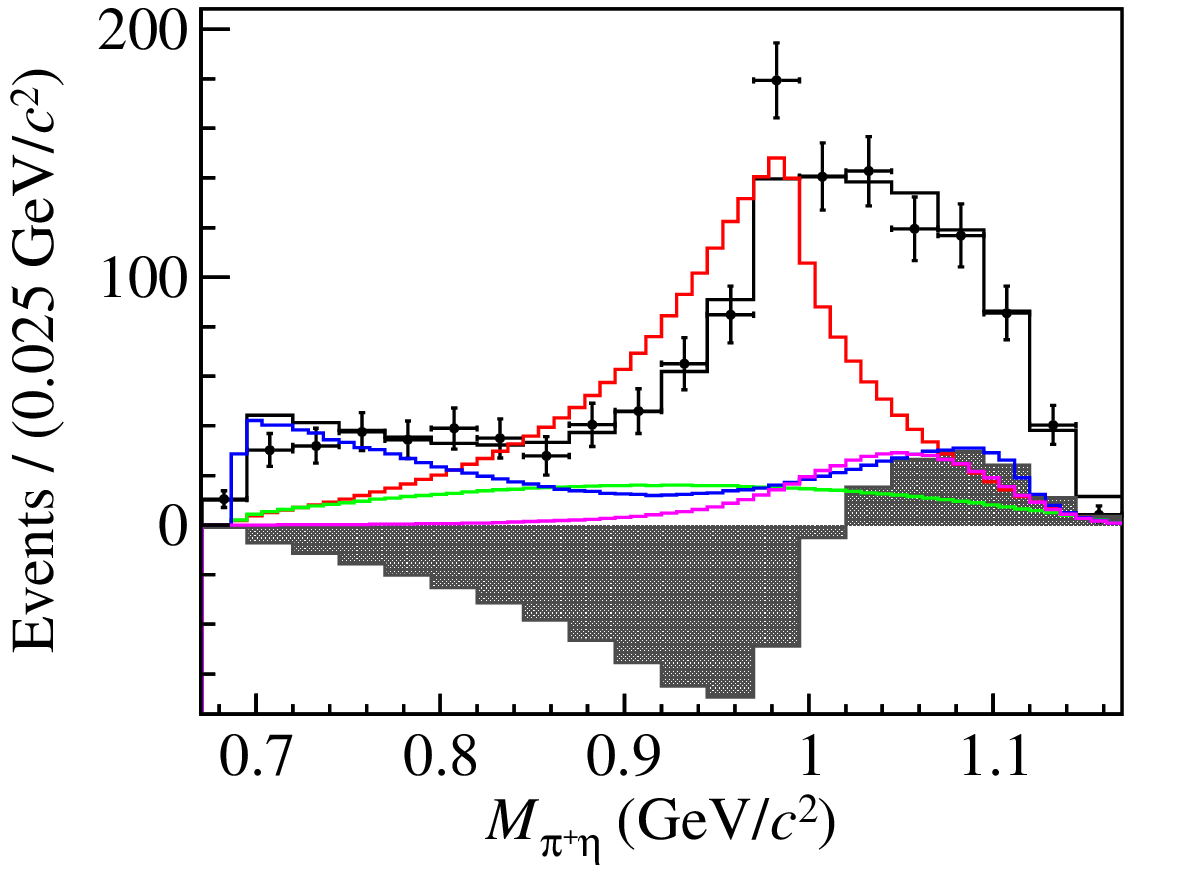}
    \includegraphics[width=0.33\linewidth]{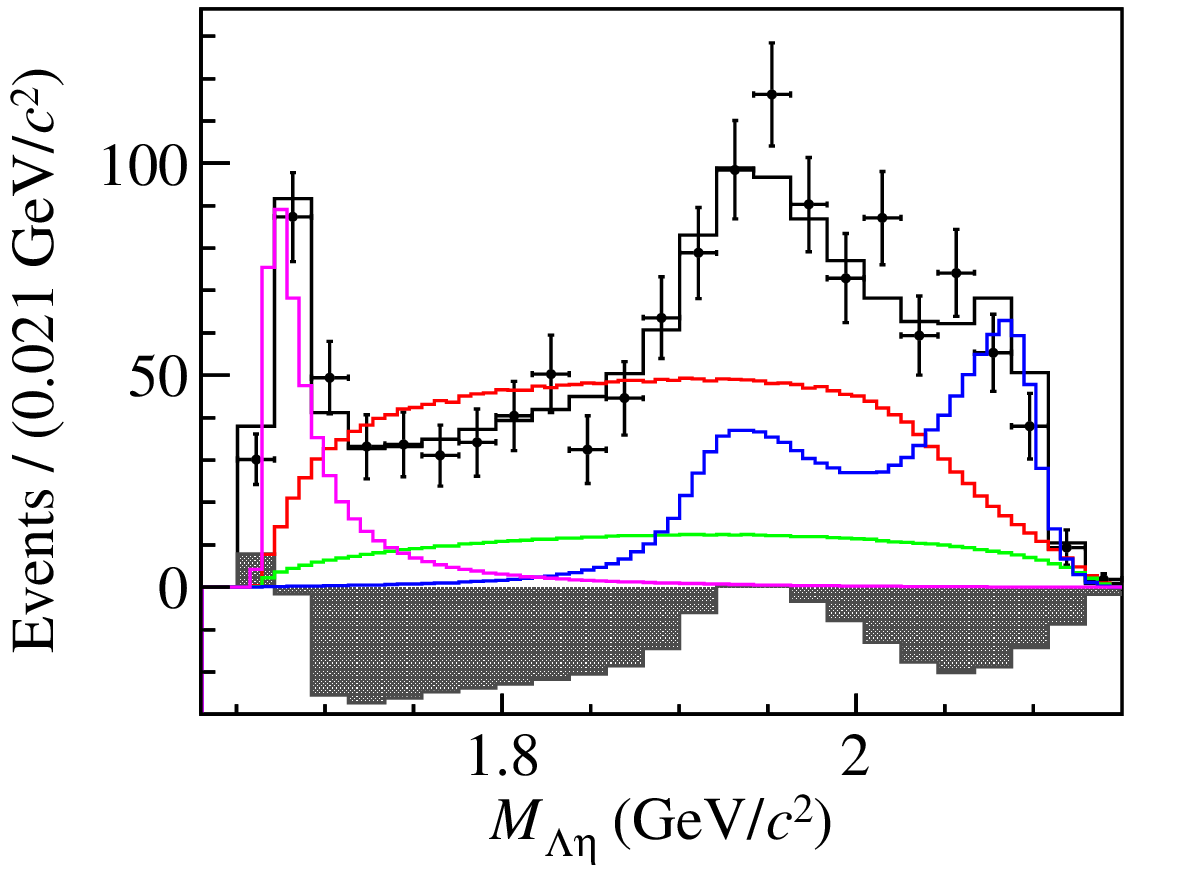}
    \caption{Top panels: projections of the fit results for $\Lambda_c^+\to\Lambda\pi^+\pi^0$ in the invariant mass spectra $M_{\Lambda\pi^+}$, $M_{\pi^+\pi^0}$ and $M_{\Lambda\pi^+}$. Bottom panels: corresponding projections for $\Lambda_c^+\to\Lambda\pi^+\eta$ in $M_{\Lambda\pi^+}$, $M_{\pi^+\eta}$ and $M_{\Lambda\eta}$ spectra.}
    \label{fig:pwa}
\end{figure}

The first partial wave analysis (PWA) of $\Lambda_c^+\to\Lambda\pi^+\pi^0$ is performed based on the helicity-amplitude formalism~\cite{BESIII:2022udq}. The fit projections onto the invariant mass spectra are shown in Fig.~\ref{fig:pwa}. The branching fractions and decay asymmetry parameters for $\Lambda_c^+\to\Sigma(1385)^+\pi^0$, $\Sigma(1385)^0\pi^+$, and $\Lambda\rho(770)^+$ are determined, as listed in Table~\ref{tab:PWA_results}. Comparison with available theoretical calculations shows that no existing model can simultaneously describe both the branching fractions and decay asymmetries.

\begin{table}[htbp]
  \centering
  \caption{Measured branching fractions and decay asymmetries of the intermediate processes in the PWA of $\Lambda_c^+\to\Lambda\pi^+\pi^0$ and $\Lambda_c^+\to\Lambda\pi^+\eta$ decays.}
  \label{tab:PWA_results}
  \begin{tabular}{c|c|c}
    \hline
    Process & $\mathcal{B}$ & $\alpha$ \\
    \hline
    $\Lambda_c^+\to\Lambda\rho$ & $(4.06\pm0.52)\times10^{-2}$ & $-0.763\pm0.070$ \\
    $\Lambda_c^+\to\Sigma(1385)^+\pi^0$ & $(5.86\pm0.80)\times10^{-3}$ & $-0.917\pm0.089$ \\
    $\Lambda_c^+\to\Sigma(1385)^0\pi^+$ & $(6.47\pm0.96)\times10^{-3}$ & $-0.79\pm0.11$ \\
    $\Lambda_c^+\to\Lambda a_0(980)^+[a_0(980)^+\to\pi^+\eta]$ & $(1.05\pm0.18)\times10^{-2}$ & $-0.91^{+0.18}_{-0.09}\pm.08$ \\
    $\Lambda_c^+\to\Sigma(1385)^+\eta$ & $(6.78\pm0.76)\times10^{-3}$ & $-0.61\pm.15\pm.04$ \\
    $\Lambda_c^+\to\Lambda(1670)\pi^+[\Lambda(1670)\to\Lambda\eta]$ & $(2.74\pm0.62)\times10^{-3}$ & $.21\pm0.27\pm0.33$ \\
    \hline
  \end{tabular}
\end{table}

\section{Partial wave analysis of $\Lambda_c^+\to\Lambda\pi^+\eta$}\label{sec:pwa_lmdpieta}

A similar PWA of $\Lambda_c^+\to\Lambda\pi^+\eta$ is performed for the first time~\cite{BESIII:2024mbf}. As shown in Fig.~\ref{fig:pwa}, the intermediate states $a_0(980)^+$, $\Sigma(1385)^+$, and $\Lambda(1670)$ are observed in the $\pi^+\eta$, $\Lambda\pi^+$, and $\Lambda\eta$ systems, respectively. The branching fractions and decay asymmetry parameters of these intermediate states are reported, as summarized in Table~\ref{tab:PWA_results}. The measured $\mathcal{B}(\Lambda_c^+\to\Lambda a_0(980)^+)$ exceeds theoretical expectations by one to two orders of magnitude~\cite{Sharma:2009zze,Yu:2020vlt}. We also test the existence of $\Sigma(1380)^+$ in a variety of alternative models, and its significance remains above $3\sigma$ in all cases, providing the first experimental evidence for this state.

\section{Summary}\label{sec:sum}

Using data samples collected near the $\Lambda_c^+\bar{\Lambda}_c^-$ production threshold, BESIII has produced a broad set of new results on $\Lambda_c^+$ decays. These include measurements of the inclusive decays $\Lambda_c^+\to Xe^+\nu_e$, $\bar{\Lambda}_c^- \to \bar{n}X$, and $\Lambda_c^+ \to K_S^0 X$; the AI-assisted observation of the semi-leptonic decay $\Lambda_c^+ \to ne^+\nu_e$; an improved measurement of the SCS decay $\Lambda_c^+ \to p\pi^0$; the first studies of $K_S^0-K_L^0$ asymmetries in $\Lambda_c^+$ decays; an angular analysis of $\Lambda_c^+\to\Xi^0K^+$; and the first amplitude analyses of $\Lambda_c^+\to\Lambda\pi^+\pi^0$ and $\Lambda_c^+\to\Lambda\pi^+\eta$. BEPCII and BESIII are currently undergoing an upgrade. The luminosity around 4.7~GeV is expected to increase by about a factor of three, and the maximum center-of-mass energy will be extended to 5.6~GeV. These improvements will enable more precise studies of $\Lambda_c^+$ decays and open new opportunities to explore other charmed baryons, such as $\Sigma_c$, $\Xi_c$, and $\Omega_c$~\cite{Li:2025nzx}.

%\section*{Acknowledgments}
%This work was supported by %the Korea National Research Foundation under grant No. 2023R1A2C300302312 and No. 2023K2A9A1A0609492411.

\end{document}